\documentclass[paper, 12pt, letterpaper, epsf]{JHEP} 
\input{epsf.tex}
\usepackage{graphics} 
\usepackage{epsfig}

\def\Bbb{\bf} 

\def\C{{\Bbb C}} 
\def\R{{\Bbb R}}
\def\Z{{\Bbb Z}} 
\def\H{{\Bbb H}} 

\def\P{{\Bbb P}}
\def\i{{\Bbb i}}\def\j{{\Bbb j}}\def\k{{\Bbb k}}

\def\id{\protect{{1 \kern-.28em {\rm l}}}}

\newcommand{\be}{\begin{equation}} \newcommand{\ee}{\end{equation}}
\newcommand{\bea}{\begin{eqnarray}} \newcommand{\eea}{\end{eqnarray}}
\newcommand{\beann}{\begin{eqnarray*}} \newcommand{\eeann}{\end{eqnarray*}}
\newcommand{\bfig}{\begin{figure}} \newcommand{\efig}{\end{figure}}

\newcommand{\ba}{\begin{array}}\newcommand{\ea}{\end{array}}

\newtheorem{Proposition}{Proposition}[section]

\newtheorem{Theorem}{Theorem}[section]
\newtheorem{Lemma}{Lemma}[section]
\newtheorem{Corrolary}{Corrolary}[section]

\newcommand{\bp}{\begin{Proposition}} \newcommand{\ep}{\end{Proposition}} 
\newcommand{\bt}{\begin{Theorem}} \newcommand{\et}{\end{Theorem}} 
\newcommand{\bl}{\begin{Lemma}} \newcommand{\el}{\end{Lemma}} 
\newcommand{\bc}{\begin{Corrolary}} \newcommand{\ec}{\end{Corrolary}} 

\title{Enhanced gauge symmetry from `toric' $G_2$ cones}

\author{L. Anguelova, C. I. Lazaroiu
\\C.~N.~Yang Institute for Theoretical Physics\\
SUNY at Stony Brook, NY, 11794-3840,
U.S.A.\\anguelov, calin @insti.physics.sunysb.edu}

\abstract{We review our
recent work on M-theory compactifications on `toric' $G_2$ cones, 
a class of models which generalize those recently considered by 
Acharya and Witten 
and lead to chiral matter in four dimensions. We explain our criteria
for identifying the gauge group content of such theories and briefly 
discuss the associated metrics. }

\preprint{YITP-SB-02-46}

\begin{document}



\vskip .3in

\section{Introduction} 

M-theory suggests phenomenologically interesting constructions
based on compactification from eleven dimensions.  
Among these, compactification on 7-dimensional spaces of $G_2$ holonomy
deserves special attention, since it leads naturally to $N=1$ supersymmetric
field theories in four dimensions. 

It is well-known, however, that compactification on {\em smooth} $G_2$
spaces has major phenomenological drawbacks, as it fails to produce
nonabelian gauge groups and chiral matter \cite{PT}. As pointed
out in \cite{Witten_anom}, this can be overcome by considering {\em
singular} $G_2$ backgrounds.  In such cases, nonabelian gauge symmetry
arises through $M2$-branes wrapping some vanishing two-cycles
\cite{Achar}, while net chirality can occur for purely topological
reasons \cite{Witten_anom}.  This mechanism was illustrated in
\cite{Witten_Acharya} by considering a specific class of conical $G_2$
spaces, some of which admit a IIA description in terms of intersecting
D6-branes.  In particular models, the appearance of chiral matter can
also be inferred directly from IIA constructions \cite{CSU1,
CSU2}. Such type II systems consist of D6-branes/orientifold 6-planes,
thus admitting a lift to an $M$-theory background of $G_2$ holonomy.

The class of spaces studied in \cite{Witten_Acharya} is obtained as
follows \cite{BS, GP}. Starting with an Einstein self-dual space $M$
(of positive scalar curvature),
one considers its six-dimensional twistor space $Y$, which can be
written as the sphere bundle associated with the vector bundle
$\Lambda^{2,-}T^*M$ of antiselfdual two forms. The twistor space
carries a Kahler-Einstein metric: 
\be
d\rho^2 = d\sigma^2+|d_A{\vec u}|^2~~,
\ee
where $d\sigma^2$ is the metric on $M$, $A$ is the connection induced
on $\Lambda^{2,-}T^*M$ by the Levi-Civita connection of $M$, and
${\vec u}=(u^1,u^2,u^3)$ are coordinates on the $S^2$ fiber, subject
to the constraint $|{\vec u}|^2=1$.  To obtain a $G_2$ space, one
takes the metric cone over $Y$, where the latter is endowed with the 
{\em modified} metric 
$d\rho'^2=\frac{1}{2}(d\sigma^2+\frac{1}{2}|d_A{\vec u}|^2)$. 
In fact, this cone admits a one-parameter family of $G_2$
deformations given by \cite{BS, GP}:
\be
\label{G2metric}
ds_{G_2}^2=\frac{1}{1-(r_0/r)^4}dr^2+\frac{r^2}{2}
(d\sigma^2+\frac{1}{2}(1-(r_0/r)^4)|d_A{\vec u}|^2)~~,~~r_0\geq 0~~.
\ee 
The conical limit is reached for $r_0=0$. It is clear from this 
expression that the
singularities of the conical $G_2$ metric are determined by the singularities
of $Y$: except for the apex, the former are obtained by taking 
the cone over the latter (and their singularity type coincides with that 
in $Y$).
The importance of studying such singularities is clear since 
they determine the essential physics of M-theory on our 
backgrounds \cite{Witten_Acharya}.

To specify the $G_2$ space completely, one must know the
(positive curvature) 
Einstein-selfdual metric on $M$.  If $M$ is smooth, then there are
only two choices, namely $S^4$ and $\C\P^2$. For these cases, M-theory
physics on the associated $G_2$ spaces was studied in
\cite{AW}. Allowing for singularities in $M$ leads to many other
examples. Among these are the models considered in
\cite{Witten_Acharya}, which correspond to $M = W\C\P^2_{p,q,r}$,
endowed with the Einstein-selfdual metric obtained implicitly in
\cite{GL}. In these examples, the singularities of the twistor space
can be determined by elementary methods \cite{Witten_Acharya}.
However, this becomes a rather difficult task if one wishes to
consider more general situations.

A characteristic feature of the examples of  \cite{Witten_Acharya} 
is the presence of a two-torus of isometries. It is by now well-known 
\cite{GL,CP, AG, BGMR, BG} that there exist infinitely
many inequivalent compact ESD orbifolds\footnote{By this we mean 
a {\em local} orbifold (V-manifold). The spaces under consideration are 
generally {\em not} global quotients of some manifold by a finite group.}  
admitting two commuting isometries. 
Such spaces provide a vast generalization of the models
discussed in \cite{Witten_Acharya}. In \cite{toric}, we developed a method
for analyzing the singularities of the twistor space 
(and thus of the associated $G_2$ cone and of its deformation 
(\ref{G2metric})) for this much larger class of `toric' ESD orbifolds. 

The approach of \cite{toric} relies on a correspondence between ESD
orbifolds and certain hyperkahler cones (a hyperkahler cone is a
hyperkahler space which is the metric cone over a compact Riemannian
space).  Recall that the ESD property of the metric amounts to the
quaternion-Kahler condition in four dimensions.  Now, it is well-known
that every quaternion-Kahler space $M$ has an associated hyperkahler
cone\footnote{The correspondence of \cite{Swann} holds irrespective of
the sign of the scalar curvature of $M$.  In the case of interest for
us (namely when $M$ has positive scalar curvature) the hyperkahler
metric on $X$ will be Riemannian (positive-definite).  Negative
curvature quaternion-Kahler spaces have pseudo-Riemannian hyperkahler
cones, a situation which is of interest in supergravity.}  $X$
\cite{Swann}, and this correspondence can be `inverted' i.e. $M$ can
be re-constructed given $X$. This allows one to translate problems in
quaternion-Kahler geometry into the language of hyperkahler spaces,
which tends to be more amenable to a solution.  In our case, the
hyperkahler cone is real eight-dimensional and will admit a $T^2$'s
worth of isometries, induced from the isometries of $M$.
$4n$-dimensional hyperkahler spaces with an $n$-torus of isometries
have been considered in the mathematics literature \cite{BD} and are
called {\it toric hyperkahler}. Our particular situation fits into
that theory for $n=2$. Such spaces can be described as hyperkahler
quotients by certain torus actions, much in the spirit of usual toric
geometry (which is concerned with {\em Kahler} toral quotients). This
allows one to reduce geometric questions for $X$ to problems in
integral linear algebra and combinatorial convex geometry.

The twistor space of $M$ can be obtained from $X$ by performing a
certain $U(1)$ {\em Kahler} quotient at a positive moment map
level. This description allows one to extract the singularities of $Y$
by first determining the singularities of $X$ and then studying the
effect of this Kahler reduction \cite{toric}.
The outcome of this analysis is a simple algorithm for identifying the
singularities of the twistor space, which we summarize in the next
section. In \cite{toric} we also used geometric arguments to describe
the type IIA reduction along one of the $U(1)$ isometries of the
associated $G_2$ space.  For a generic choice of the reduction
isometry, the resulting IIA background is strongly coupled, and does
not seem to admit a simple perturbative description. However, certain
models admit a `good isometry', which leads to a system of $D6$-branes
upon reduction.  Such an isometry (when it exists) can be found by a
simple criterion discussed in \cite{toric}.  Since the IIA metric
inherits a $U(1)$ isometry induced from the M-theory solution, one can
T-dualize along its direction to obtain a IIB background.  This will
correspond to a system of delocalized 5-branes if a `good isometry'
is used for IIA reduction.  While models with a good isometry admit a
simple IIA/IIB description, we stress that there is no obvious reason
to discard other models. From the M-theory perspective, models without
a good isometry are as good as any other.

The abstract arguments of \cite{toric} were confirmed in \cite{metrics} 
by direct calculation. Using the result of \cite{CP}, we wrote
down the $G_2$ metric obtained from a general ESD space of positive
scalar curvature and admitting a $T^2$ of isometries.  By performing
the reduction, we found the corresponding T-dual IIA and IIB
backgrounds. Guided by the analysis of \cite{toric} and the
topological arguments of \cite{BGMR}, we computed the explicit
asymptotics of relevant fields (metric, coupling constant, and
R-R/NS-NS forms) in the vicinity of the branes and determined the
relevant RR fluxes. The main results of \cite{toric} and
\cite{metrics} are briefly reviewed below.

\section{Singularities of the twistor space}

The toric hyperkahler cones associated with our ESD spaces are
obtained as hyperkahler quotients of some affine quaternion space
$\H^n$ by an appropriate torus action. Namely, we have 
$X=\H^n///_0U(1)^{n-2}$, where
the subscript $0$ indicates that the hyperkahler quotient is taken
at zero moment map levels. The action of $U(1)^{n-2}$ on the quaternion
coordinates $u_1\dots u_n\in \H$ is given by: \be
\label{action}
u_k \rightarrow
\prod_{\alpha=1}^{n-2}{\lambda_\alpha^{q_k^{(\alpha)}}}u_k~~, \ee
where $\lambda_{\alpha}$ are complex numbers of unit modulus. The
$(n-2)\times n$ matrix\footnote{One has to impose some 
mild conditions on $Q$
to guarantee that the $U(1)^{n-2}$ action on $\H^n$ is
effective. See \cite{toric} and footnote 6 of this letter.}
$Q_{\alpha k}=q_k^{(\alpha)}$ determines the cone $X$.
Following toric geometry procedure, we introduce a 
$2\times n$ matrix $G$ whose  
rows form an integral basis for the kernel
of $Q$. The columns of $G$ are two-dimensional integral vectors $\nu_1
\dots \nu_n$, which we call {\it toric hyperkahler
generators}. Splitting every quaternion coordinate into its complex
components: \be u_k = w_k^{(+)} + \j w_k^{(-)} ~~, \ee (where
$\i,\j,\k$ are the imaginary quaternion units) we obtain 
the following re-writing of the torus action: \be
\label{complex_action}
w_k^{(+)}\rightarrow \prod_{\alpha=1}^{n-2} \lambda_\alpha
^{q^{(\alpha)}_k}w_k^{(+)}~~,~~ w_k^{(-)}\rightarrow
\prod_{\alpha=1}^{n-2} \lambda_\alpha ^{-q^{(\alpha)}_k}w_k^{(-)}~~.
\ee The twistor space is obtained as a $U(1)$ Kahler quotient 
$Y=X//_{\zeta}U(1)$  of $X$ at a
positive level $\zeta$ (the precise choice of $\zeta$ 
fixes the overall scale of $Y$ and $M$). The $U(1)$ action is given by: 
\be
\label{U1proj}
w_k^{(\pm)}\rightarrow \lambda w_k^{(\pm)}~~  
\ee
and will be called `the projectivizing action'. For simplicity, 
we shall assume
\footnote{If this assumption does not hold, then the projectivizing 
$U(1)$ has a trivially acting $\Z_2$ subgroup. In this case, there 
are {\em two} toric hyperkahler cones associated with $Y$ and 
the results described below must be slightly modified. The required 
modifications are described in \cite{toric}.}
that this action is effective on $X$. 

In \cite{toric}, we show that the singular locus of $Y$ is a subset 
of the so-called 
{\it distinguished locus} $Y_D$. The latter consists of two kinds of
holomorphically embedded two-spheres, which are distinguished 
by their position with respect to the $S^2$ fibration $Y\rightarrow M$:

1) $Y_D$ contains $n$ {\it vertical} spheres $Y_j$ ($j=1\dots n$), 
which are fibers of $Y$ over $M$.

2) it also contains $2n$ {\it horizontal} spheres 
$Y_e$, which are lifts of spheres lying in $M$. 

\noindent The meaning of the index $e$ will become apparent in a 
moment (it corresponds to the edges of a certain polygon). The union 
$Y_V$ of all vertical spheres and the union $Y_H$ of horizontal spheres 
will be called the {\em vertical} and {\em horizontal} loci.

A basic observation (which goes back to \cite{HKLR}, see also
\cite{BD}) is that the hyperkahler moment map constraints are solved
by points $u\in \H^n$ whose complex coordinates are solutions of the
system: \be
\label{ab}
\frac{1}{2}(|w^{(+)}_k|^2-|w^{(-)}_k|^2)=\nu_k\cdot a~~,~~
w^{(+)}_kw^{(-)}_k=\nu_k\cdot b~~, ~{\rm~for~all~}k=1\dots n~~,
\label{X} \ee for some parameters $a=(x_1, y_1)\in \R^2$ and $b=(x_3+ix_2,
y_3+iy_2)\in \C^2$. Using this, one shows \cite{toric}
that the distinguished locus can be
described in terms of the {\it characteristic polygon}: \be
\label{polygon} \Delta=\{a\in \R^2|\sum_{k=1}^n{|\nu_k\cdot
a|}=\zeta\}~~.  \ee 
This is a planar convex polygon with $2n$ vertices, which is symmetric
under the point reflection $a \rightarrow -a$.  In particular,
$\Delta$ has $n$ {\em principal diagonals} $D_j=\{a\in \R^2|a\cdot
\nu_j=0\}$. These are the diagonals which pass through the origin and
thus connect opposite vertices.

\begin{figure}[hbtp]
\begin{center}
\scalebox{0.6}{\input{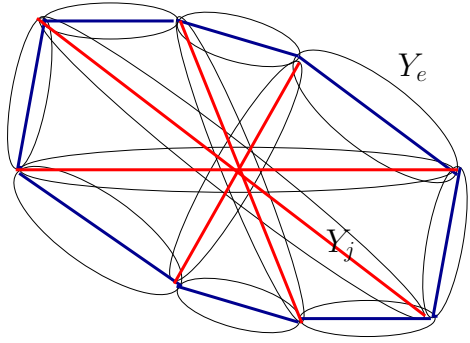}}
\end{center}
\caption{\label{dist} Fibration of the distinguished locus over the 
characteristic polygon 
$\Delta$. Since we attempt to represent this in two dimensions, it may 
appear that some spheres intersect at more than isolated points. This is 
{\em not} actually the case. }
\end{figure}

Each vertical sphere $Y_j$ is defined by the condition $u_j = 0$
(i.e. $w_j^{(+)} = w_j^{(-)} = 0$) with $j=1\dots n$. The horizontal
spheres $Y_e$ are indexed by the edges $e$ of $\Delta$ and are
determined as follows. Given such an edge, define the signs
$\epsilon_j(e) = sign(\nu_j\cdot p_e)$, where $p_e$ is the two-vector
from the origin of the plane to the middle point of $e$.  Then $Y_e$
is the locus in $Y$ given by the equations $w_1^{(-\epsilon_1)} =
w_2^{(-\epsilon_2)} = \dots = w_n^{(-\epsilon_n)} = 0$. As discussed
in \cite{toric},  $Y_j$ is a circle fibration over
the principal diagonal $D_j$ (the circle fibers collapse to points
above the two vertices connected by the diagonal). Each horizontal
sphere $Y_e$ is a circle fibration over the edge $e$, whose circle
fibers collapse to points above the vertices connected by this
edge. The twistor space admits an antiholomorphic involution acting
along its $S^2$ fibers. Its restriction to $Y_H$ intertwines the
horizontal spheres $Y_e$ and $Y_{-e}$, while covering the point
reflection $\iota :a\rightarrow -a$ of $\Delta$. Finally, the two
horizontal spheres associated with adjacent edges of $\Delta$ and the
vertical sphere associated with the principal diagonal passing through
their common vertex have precisely one point in common. Such {\em
special points} of the distinguished locus are in one-one
correspondence with the vertices of $\Delta$. This situation is
summarized in figure \ref{dist}.

As mentioned above, the singular locus of the twistor space is a 
subset of its distinguished locus. Namely, some of the vertical and 
horizontal spheres will consist of cyclic singularities, 
while some of the special points will carry singularities of their own. 
The relevant singularity types are described by the following rules:

(a) Given a horizontal sphere $Y_e$, consider the integral vector
\footnote{With our assumptions on $Q$, this vector can never 
vanish.}: 
\be
\label{nu} \nu_e=\sum_{k=1}^n{\epsilon_k(e)\nu_k}~~, 
\ee 
where the signs $\{\epsilon_k(e)\}$ are defined as explained above.
Then $Y$
has a $\Gamma_e=\Z_{m_e}$ quotient singularity along $Y_e$, where 
$m_e$ is the greatest common divisor of the two entries of the vector 
$\nu_e$.  
The orbifold action on the coordinates transverse to $Y_e$ can 
be determined as explained in \cite{toric}. 

(b) Given a vertical sphere $Y_j$, consider the matrix ${\tilde Q}_j$
obtained by deleting the $j^{th}$ and $(j+n)^{th}$ columns of the
$(n-1)\times (2n)$ matrix ${\tilde Q} =
\left[\begin{array}{ccc}Q&~&-Q\\1&\dots&~1\end{array}\right]$.  Then
the singularity group $\Gamma_j$ of $Y$ along $Y_j$ coincides with
$\Z_{m_j}$ or $\Z_{2m_j}$, where $m_j$ is the greatest common divisor
of the components of $\nu_j$.  
To find which of these cases occurs,
one computes the integral Smith form\footnote{Given an integral
$r\times n$ matrix $F$ with $r\leq n$, one can find matrices $U\in
GL(r,\Z)$ and $V\in GL(n,\Z)$ such that the matrix
$F^{ismith}=U^{-1}FV$ (the {\em integral Smith form} of $F$) has
the form $[D,0]$. Here $D=diag(t_1 \dots t_r)$ with $t_1 \dots
t_r$ some non-negative integers satisfying the division relations
$t_1|t_2|\dots |t_r$. These integers are called the {\em invariant
factors} of $F$. The mild condition 
mentioned in footnote 3 is that 
all invariant factors of $Q$ be equal to $1$.} 
of the matrix ${\tilde Q}_j$: \be {\tilde Q}_j^{ismith}=[diag(1\dots 1,
t_j),0]~~, \ee where $t_j=m_j$ or $t_j=2m_j$. The singularity group
$\Gamma_j$ coincides with $\Z_{t_j}$. 
The orbifold action on the transverse coordinates can be determined 
as explained in \cite{toric}.

Note that some of the vertical and horizontal spheres will be smooth 
(this happens respectively when $m_e=1$ and $t_j=1$).
The singularity type at the special points (which correspond to the vertices of $\Delta$) can also be determined 
by simple criteria, which the interested reader 
can find in \cite{toric}.

The criteria listed above determine the 
gauge groups produced at the singular loci of the $G_2$ cone: 
a $\Z_m$ singularity corresponds to an $SU(m)$ gauge factor. 
On the other hand, the second Betti number of our $G_2$ cone equals $n-1$. 
As in \cite{Witten_Acharya},  M-theory on such cones
produces an Abelian $U(1)^{n-1}$ factor, and 
chiral fermions charged under this group and 
localized at the cones' apex. The second Betti number reduces to $n-2$
(and the Abelian gauge factor is Higgsed 
down to $U(1)^{n-2}$) when deforming the cone to the solution with 
non-vanishing $r_0$. All of this is qualitatively identical with the 
behavior of the models discussed in \cite{Witten_Acharya}.

Dimensionally reducing M-theory through one of the $T^2$ isometries, 
one obtains a type IIA background with a non-vanishing RR one-form. 
Such an isometry is called {\em good} 
if its fixed point set coincides with the singular locus of the
$G_2$ cone. If a good isometry exists, then the associated type IIA 
reduction gives a system
of D6-branes. It is shown in \cite{toric} that a 
good isometry exists if all two-vectors $\nu_e,\nu_j$ are collinear; 
in this case, the good isometry corresponds to their common direction, 
viewed as a direction in the Lie algebra of the two-torus of isometries 
of the $G_2$ space.

Since the resulting IIA background inherits a 
$U(1)$ isometry, one can T-dualize to a IIB solution. 
It turns out that the orbits of the T-dualizing isometry
lie along the $S^1$ fibers of the loci $Y_j\rightarrow D_j$ 
and $Y_e\rightarrow e$, both of which descend to the IIA solution. 
In the good isometry 
case, it follows that T-duality acts along the worldvolume directions, 
thereby producing delocalized 5-branes in type IIB.
This abstract argument can be substantiated by a direct analysis of 
the relevant field configurations, which was carried out in \cite{metrics}
and is briefly reviewed below.  

\section{$G_2$ metrics and reduction to type II string theory}

As discussed above, the ESD space $M$ and its associated 
$G_2$ cone, as well as the hyperkahler 
cone $X$ admit a two-torus of isometries for 
our class of models. The most general ESD metric admitting two commuting 
isometries was given explicitly in recent work of Calderbank and Pedersen 
\cite{CP}. To describe the solution, one introduces 
coordinates $\phi$, $\psi$ along  the $T^2$ fibers of $M$ and 
$\rho, \eta$ along the base of its $T^2$ 
fibration. There exists an $SL(2,\Z)$ freedom in the choice of $\phi$ and 
$\psi$, which is related to modular transformations of the lattice $\Z^2$
containing the toric hyperkahler generators.
As explained in \cite{CP}, the base coordinates can be chosen 
such that $\rho\geq 0$ and
$\eta\in \R$, and parameterize the upper half plane (the hyperbolic
plane ${\cal H}^2$). The metric of \cite{CP} takes the form:

{\footnotesize \bea \!\!\!\!\!\!\!\!\!\!  d\sigma^2=\frac{ F^2 -
4\rho^2(F_\rho^2 + F_\eta^2) }{4 F^2}\; \frac{d\rho^2 +
d\eta^2}{\rho^2} +\frac{ \left[ (F - 2 \rho F_\rho) \alpha - 2 \rho
F_\eta \beta \right]^2 + \left[ -2\rho F_\eta \alpha + (F + 2 \rho
F_\rho )\beta \right]^2 }{ F^2\left[F^2-4\rho^2(F_\rho^2 +
F_\eta^2)\right]}, \label{Mmetric} \eea} where \be
\label{F}
F=\sum_{k=1}^n \frac{\sqrt{(\nu_k^2)^2\rho^2
+(\nu_k^2\eta+\nu_k^1)^2}}{\sqrt\rho} \ee and
$\alpha=\sqrt\rho\,d\phi$, \,\, $\beta=(d\psi+\eta\,
d\phi)/\sqrt\rho$, and $F_{\rho}= \partial F/\partial \rho$, $F_\eta=
\partial F/\partial \eta$. The
metric on the $S^2$ fiber of the twistor space of $M$ takes the form:
\be \label{fmetric} |d_A {\vec u} |^2=(du^i + \epsilon^{ijk} A^j
u^k)^2~~, \ee where ${\vec u}=(u^1,u^2,u^3)$ are constrained 
coordinates $(\sum_{i=1}^3
(u^i)^2 = 1)$ on the fiber and the connection $A$ is given by 
\cite{CP}: \be A^1 = -
\frac{F_{\eta}}{F} d\rho + \left(\frac{1}{2 \rho} + \frac{F_{\rho}}{F}
\right) d\eta~~,~~A^2 = - \frac{\sqrt{\rho}}{F} d\phi~~,~~A^3 =
\frac{\eta}{F\sqrt{\rho}} d\phi + \frac{1}{F \sqrt{\rho}} d\psi \, .
\ee Substituting (\ref{Mmetric}) and (\ref{fmetric}) into
(\ref{G2metric}) we obtain the $G_2$ metric of our models.
As expected, this leads to rather complicated formulae, which 
can be found in \cite{metrics} and will not be reproduced here. 

\begin{figure}[hbtp]
\begin{center}
\scalebox{0.6}{\input{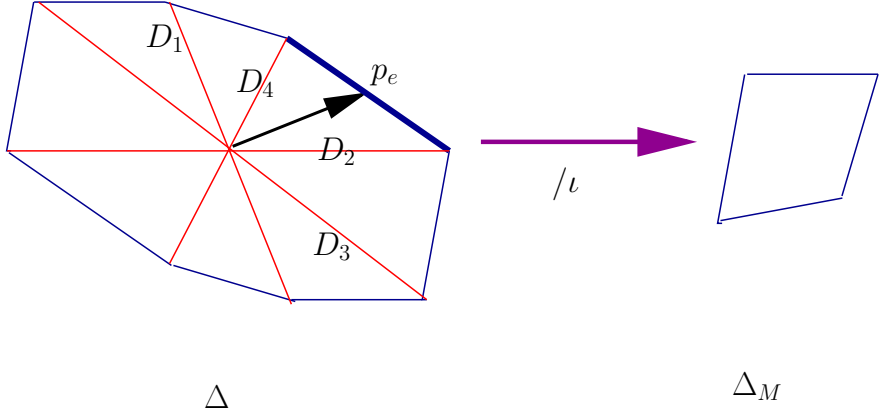}}
\end{center}
\caption{\label{DDM} Examples of the 
polygons $\Delta$ and $\Delta_M$ for $n=4$. For the edge $e$ of $\Delta$ 
drawn as a bold line, we show the vector $p_e$ used in the definition of
the signs $\epsilon_j(e)$. Note that the principal diagonals 
$D_j$ need not lie in trigonometric order.}
\end{figure}

As explained in \cite{metrics}, the 
singular loci of $Y$ project onto the boundary of 
${\cal H}^2$. The latter is obtained by adding the point at infinity
to the upper half plane: this results in 
a disk whose boundary corresponds to $\rho = 0$. In agreement with the
topological analysis of \cite{BGMR}, the vertical loci project to
points $P_1\dots P_n$ lying on this boundary, while the horizontal loci
project to the segments $P_k P_{k+1}$ connecting them.
The points $P_k$ correspond to $\rho=0$ and $\eta=\eta_k$, 
where $\eta_k=-\frac{\nu_k^1}{\nu_k^2}$.
The circular polygon (with vertices $P_k$) obtained in this manner 
can be identified topologically with the polygon $\Delta_M$
defined as the quotient of the characteristic polygon 
$\Delta$ (see (\ref{polygon})) through the 
point reflection $\iota : a \rightarrow -\,a$ of the plane: \be
\Delta_M=\Delta/\iota~~.  \ee 
The polygon $\Delta_M$ has $n$ vertices. Opposite edges $e$ and $-e$ 
of $\Delta$ cover the same edge of $\Delta_M$ (figure \ref{DDM}). 

Starting with the purely geometric background $ds_{11}^2 =
ds^2(\R^{3,1}) + ds_{G_2}^2$ of 11d supergravity (see
(\ref{G2metric})), one can reduce along one of the $T^2$ circles 
to obtain a IIA solution with non-constant dilaton and non-vanishing 
RR 1-form. Modulo a common modular transformation of the vectors $\nu_j$, 
it suffices to consider the reduction along the $\phi$-circle. 
The explicit form of the resulting IIA  solution 
can be found in \cite{metrics} and will not be reproduced here.

Parameterizing the $S^2$ fibers of $Y$ by spherical
coordinates: \be u^3=\cos\theta \, , \qquad u^1=\sin\theta \cos\chi \,
, \qquad u^2=\sin\theta \sin \chi
 \label{sphcoord} \ee 
 with $\theta \in [0, \pi] \, , \, \chi \in [0, 2 \pi]$, one obtains
coordinates $(r, \rho, \eta, \chi, \theta,\psi)$ for the internal
6-dimensional part of the resulting IIA background. For every fixed
value of $r$, one has a compact 5-manifold ${\cal N}^5$ with
coordinates $(\rho,\eta,\chi,\theta,\psi)$. This is the $U(1)_\phi$
reduction of the twistor space $Y$. Under this reduction, the
distinguished locus of the twistor space corresponds to the locus
${\cal N}_D$ in ${\cal N}^5$ defined by $\rho=0, \theta=0,\pi$ or
$\rho=0,\eta=\eta_k$ (for some $k$).  In the limit $\rho \rightarrow
0$, the angular part of the metric turns out to depend only on
$\theta$ and $\xi \equiv \nu_{\epsilon}^1 \chi + \psi$, where
$\nu_{\epsilon}^1$ is the first component of a two-component vector
$\nu_{\epsilon}$ related to the edge of $\Delta_M$ covered by the
edges $e, -e$ of $\Delta$ ($\nu_{\pm e}$ of (\ref{nu}) equals $\pm
\nu_{\epsilon}$). Hence the locus $\rho=0$ in ${\cal N}^5$ is only
three-dimensional, while the sublocus ${\cal N}_D$ has dimension
two. The horizontal part $\rho=0,\theta=0,\pi$ of ${\cal N}_D$ turns
out to be a union of $2n$ components ${\cal N}_e$ indexed by the edges
of $\Delta$. The vertical part consists of the
components ${\cal N}_k$ defined by $\rho=0$ and $\eta=\eta_k$.

The $\rho\rightarrow 0$ asymptotics of the IIA fields (dilaton, RR 1-form and
metric) was analyzed in detail in \cite{metrics}. 
Fixing a component ${\cal N}_e$ or ${\cal N}_j$, the IIA background turns out
to be strongly coupled in its vicinity unless $\nu_e^1=0$ (resp. $\nu_j^1=0$), 
i.e. unless the direction $\phi$ used for reduction corresponds to the 
direction $\nu_e$ (resp. $\nu_j$) in the Lie algebra $\R^2$ of the
$T^2$ isometry group. Remember from the previous
section that ${\cal N}_e$ (resp. ${\cal N}_j$) corresponds to enhanced
gauge symmetry if $\nu_e$ (resp. $\nu_j$) is not primitive.  To insure
weak coupling along all such loci, one must be able to choose
$U(1)_\phi$ in the direction of all non-primitive $\nu_e$ and $\nu_j$.
This requires that all these vectors be proportional (over the reals),
which is the condition for a `good isometry' mentioned above. 

\begin{figure}[hbtp]
\begin{center}
\mbox{\epsfxsize=10truecm \epsffile{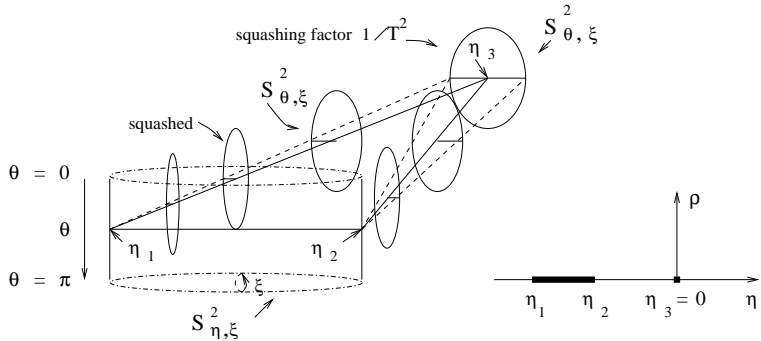}}
\end{center}
\caption{\label{2Aboundary} The locus $\rho=0$ (the 3-dimensional
boundary of the space parameterized by $\rho, \eta, \psi, \theta,
\chi$) at a fixed value of $r$.  The figure shows the case $n=3$, for
a model admitting a good isometry.  $\Delta_M$ is the triangle with
vertices $\eta_1,\eta_2,\eta_3\in \partial{\overline {\cal
H}}^2=\R\cup \{\infty\}$; these 3 points are shown in the upper half
plane on the right figure.  The picture corresponds to a good isometry
given by $\nu_\epsilon^1=0$ along the edge $[\eta_1,\eta_2]$ (singular
horizontal locus giving a pair of D6-branes at $\theta = 0, \pi$ with
worldvolume containing $S_{\eta \xi}^2$) and $\nu_3^1=0$ at the vertex
$\eta_3$ (singular vertical locus again giving a D6-brane containing
$S_{\theta \xi}^2$).}
\end{figure}

For simplicity, we shall restrict to the case when $U(1)_\phi$ is good.
In this situation, all interesting components of the locus ${\cal
N}_D$ are weakly coupled. Moreover, the computations of \cite{metrics}
show that the interesting loci ${\cal N}_e$, ${\cal N}_j$ carry
$|\nu_{e}^2|$ resp. $|\nu_j^2|$ units of RR flux.  Since
$\nu_e^1=\nu_j^1=0$, one has $|\nu_{e}^2|=m_e$ and $|\nu_j^2|=m_j$,
where $m_e=gcd(\nu_e^1,\nu_e^2)$ and $m_j=gcd(\nu_j^1,\nu_j^2)$.  This
confirms the prediction of \cite{toric} reviewed in the previous
section, and shows that the appearance of enhanced $SU(m_e)$ or
$SU(m_j)$ symmetry at these loci is due to the presence of coincident
D6-branes\footnote{The equivalence of the two mechanisms of
enhancement of gauge symmetry - coinciding D-branes in type IIA string
theory and M2-branes wrapping collapsing two-cycles at an $A-D-E$
singularity - was first pointed out in \cite{Sen}.}. 
The geometry at $\rho=0$ is depicted in
figure \ref{2Aboundary} for a model with $n=3$ which admits a good
isometry (for example the $W_{p,p,q}$ models of \cite{Witten_Acharya}).

T-dualizing along the remaining $U(1)$ isometry (parameterized by
$\psi$) one finds that the $\chi$-circle is always
collapsed to a point in the limit $\rho \rightarrow 0$ \cite{metrics}. 
Analyzing the asymptotic behavior of the IIB fields enables the computation
of the relevant NS-NS/R-R fluxes. 
When the $U(1)$ isometry along $\phi$ is good, one 
finds that the corresponding IIB loci support delocalized
five-branes, once again confirming the predictions of \cite{toric}.

\acknowledgments{
The present work was supported by the Research Foundation under NSF 
grant PHY-0098527.}


\begin{thebibliography}{150}
\bibitem{PT}{G.~Papadopoulos, P.~Townsend, {\em
Compactification of $d=11$ Supergravity on Manifolds of Exceptional
Holonomy}, Phys. Lett. {\bf B357} (1995) 300} 
\bibitem{Achar}{B.~Acharya, {\em M theory, Joyce Orbifolds and Super
Yang-Mills}, Adv. Theor. Math. Phys. 3 (1999) 227, hep-th/9812205}
\bibitem{Witten_anom}{E.~Witten, {\em Anomaly Cancellation On
Manifolds Of $G_2$ Holonomy}, hep-th/0108165.}
\bibitem{Witten_Acharya}{ B.~Acharya, E.~Witten, {\em Chiral Fermions
from Manifolds of $G_2$ Holonomy}, hep-th/0109152.}
\bibitem{CSU1}{M.~Cvetic, G.~Shiu and A.~M.~Uranga, {\em Chiral
four-dimensional $N = 1$ supersymmetric type IIA orientifolds from
intersecting D6-branes}, Nucl.Phys. {\bf B615} (2001) 3-32,
hep-th/0107166.}  
\bibitem{CSU2}{M.~Cvetic, G.~Shiu and A.~M.~Uranga,
{\em Three-Family Supersymmetric Standard-like Models from
Intersecting Brane Worlds}, Phys.Rev.Lett. {\bf 87} (2001) 201801,
hep-th/0107143.}  
\bibitem{BS}{R. L. Bryant, S. M. Salamon, {\em On
the construction of some complete metrics with exceptional holonomy},
Duke Math. J. {\bf 58} (1989)3, 829.}  
\bibitem{GP}{G.~W.~Gibbons,
D.~N.~Page, C.~N.~Pope, {\em Einstein metrics on $S^3$, $\R^3$ and
$\R^4$ bundles}, Commun. Math. Phys {\bf 127} (1990), 529-553.}
\bibitem{AW}{M.~Atiyah, E.~Witten, {\em M-theory dynamics on a
manifold of $G_2$ holonomy}, hep-th/0107177.}
\bibitem{GL}{K. Galicki, H. B. Lawson, Jr, {\em Quaternionic reduction
and quaternionic orbifolds}, Math. Ann. {\bf 282} (1988) 1-21.}
\bibitem{CP}{D. M. J. Calderbank, H. Pedersen, {\em Selfdual Einstein
metrics with torus symmetry}, math-DG/0105263.}
\bibitem{AG}{V.~Apostolov, P.~Gauduchon, {\em
Self-dual Einstein Hermitian four manifolds}, math.DG/0003162.}
\bibitem{BGMR}{C.~P.~Boyer, K.~Galicki, B.~M.~Mann, E.~G.~Rees, {\em
Compact $3$-Sasakian $7$-manifolds with arbitrary second Betti number}
Invent. Math.  {\bf 131} (1998), no. 2, 321--344.}  
\bibitem{BG}{
C.~P.~Boyer, K.~Galicki, {\em 3-Sasakian Manifolds}, Surveys in
differential geometry: Essays on Einstein manifolds, 123--184,
Surv. Differ. Geom., VI, Int. Press, Boston, MA, 1999,
hep-th/9810250.}  
\bibitem{Swann}{A. Swann, {\em Hyperk\"ahler and
quaternionic K\"ahler geometry}, Math. Ann. {\bf 289} (1991) 421.}
\bibitem{toric}{L.~Anguelova, C.~I.~Lazaroiu, {\em M-theory
compactifications on certain `toric' cones of $G_2$ holonomy},
hep-th/0204249.}  
\bibitem{metrics}{L. Anguelova, C. I. Lazaroiu, {\em
M-theory on `toric' $G_2$ cones and its type II reduction},
hep-th/0205070.}  
\bibitem{Sen}{A.~Sen, {\em A note on enhanced gauge
symmetries in M- and string theory}, JHEP 9709 (1997) 001,
hep-th/9707123} 
\bibitem{HKLR}{N. Hitchin, A. Karlhede, U. Lindstr\"{o}m,
M. Ro\v{c}ek, {\em Hyperk\"{a}hler Metrics and Supersymmetry}, 
Commun. Math. Phys. {\bf 108} (1987) 535.}
\bibitem{BD}{R. Bielawski, A. S. Dancer, {\em The
geometry and topology of toric hyperkahler manifolds},
Commun. Anal. Geom, vol 8, No. 4 (2000), 727-759.}
\end{thebibliography}
\end{document}